\documentclass{article}

\usepackage{arxiv}

\usepackage{amssymb,amsmath}
\usepackage[utf8]{inputenc} 
\usepackage[T1]{fontenc}    
\usepackage{hyperref}       
\usepackage{url}            
\usepackage{booktabs}       
\usepackage{amsfonts}       
\usepackage{nicefrac}       
\usepackage{microtype}      
\usepackage{lipsum}		
\usepackage{graphicx}
\usepackage{natbib}
\usepackage{doi}

\usepackage{subcaption}
\usepackage{wrapfig}

\DeclareMathOperator*{\argmax}{arg\,max}

\begin{document}

\title{Matching Theory-based Recommender Systems in Online Dating}

\author{
    Yoji Tomita\\
    CyberAgent, Inc.\\
    \texttt{tomita\_yoji@cyberagent.co.jp}
    \And
    Riku Togashi\\
    CyberAgent, Inc.\\
    \texttt{togashi\_riku@cyberagent.co.jp}
    \And
    Daisuke Moriwaki\\
    CyberAgent, Inc.\\
    \texttt{moriwaki\_daisuke@cyberagent.co.jp}
}

\date{}




\maketitle


\everypar{\looseness=-1}

\section{Introduction}
\label{sec:intro}

Online dating platforms provide people with the opportunity to find a partner.
Recommender systems in online dating platforms suggest one side of users to the other side of users.
This \emph{reciprocal recommendation problem} arises in online dating \citep{Pizzato2010-dm, Xia2015-nf}, job recommendation \citep{Almalis2014-md} and peer learning process \citep{Potts2018-uc}.

\emph{Reciprocal recommender systems (RRSs)} must take into account two aspects which do not arise in standard recommender systems. One aspect is the mutual interests of users.
Even if one user has a strong interest in a particular user, she/he may have no interest in him/her at all.
Thus, RRSs should be based on the interests of both sides of users.
The other important aspect is \emph{capacities} of the users; it is impossible for a user to keep up with all the candidates.
Popular users are often recommended very frequently. As a result, a few \emph{super stars} receive a large proportion of \textit{likes}, overwhelming the time that they can spend for screening.
RRSs should be designed to alleviate such problems to increase user satisfaction.

The two-sided matching problem is to ``match'' one side of people and the other side in a situation where each person has a different preference and a matching capacity.
This field is called \emph{matching theory}, and various algorithms have been developed since \citet{Gale1962-gc, Shapley1971-fg}.
Matching theory is applied in the analyses of marriage markets \citep{Becker1973-jq, Choo2006-kx}, school choice \citep{Abdulkadiroglu2003-au} and job matching for physicians \citep{Roth1999-zy}.

In this talk, we discuss the potential interactions between RRSs and matching theory.
We also present our ongoing project to deploy a \emph{matching theory-based recommender system (MTRS)} in a real-world online dating platform. This talk covers other important directions regarding RRSs, including scalability, algorithmic fairness, bandit algorithm, and online experimentation.

\section{Reciprocal Recommendation} \label{sec:recipro}
In this section, we present a brief introduction of RRSs.
In particular, we focus on fusion approaches and aggregate functions to compute reciprocal preference scores that represent the degree of mutual preference.

In a RRS for online dating, unilateral preference scores $p_{x, y} \in [0, 1]$, representing the preference of man $x \in X$ for woman $y \in Y$, and the opposite $p_{y, x} \in [0, 1]$ are predicted as in standard recommender systems, where $X$ and $Y$ are the set of men and women, respectively.\footnote{In this paper, we assume a heterosexual matching platform because our application only considers such matching. However, the above discussion can be extended to other types of matching.}
Each of these scores is predicted by content-based \citep{Pizzato2010-dm} or collaborative filtering-based \citep{Xia2015-nf, Neve2019-fe} algorithms.
To take into account the ``reciprocity'', bilateral preference scores $p_{x \leftrightarrow y}$ are calculated using some fusion approaches.
The most common fusion approach is to aggregate unilateral preference scores: $p_{x \leftrightarrow y} = \phi(p_{x, y}, p_{y, x})$ with some aggregate function $\phi:[0,1]\times[0,1] \to [0, 1]$.

Various implementations of the aggregation function have been explored, e.g., harmonic mean~\citep{Pizzato2010-dm, Xia2015-nf}, arithmetic mean~\citep{Neve2019-fe, Neve2019-fa}, geometric mean~\citep{Neve2019-fe, Neve2019-fa}, cross-ratio uniform~\citep{Appel2017-gn}, matrix multiplication \citep{Jacobsen2019-vl}, weighted mean with optimized weighting parameters \citep{Kleinerman2018-fc} and multiplicative inverse of rank multiplication \citep{Mine2013-ab}.

Although several fusion approaches are used in studies and practice, fusion approaches have been relatively less analyzed.
Most of them try to capture the reciprocity of preferences but rarely take into account the capacities of the users. The introduction of matching theory in RRS would be a promising treatment of this point.

\section{Matching with transferable utility} \label{sec:matching}
We briefly introduce the \citet{Choo2006-kx} model, which extends \emph{matching with transferable utility (TU matching)} \citep{Shapley1971-fg, Becker1973-jq}.
TU matching model assumes that two types of agent choose their partner in the decentralized market, and monetary transfer occurs between agents who are matched.
Transfers can be interpreted as, for example, wages in job matching markets, share of household economy in marriage markets, or gifts in dating markets.
We consider equilibrium matching, in which the demands of both sides coincide by adjusting the transfer amount, just as prices in a market economy.
TU matching model has been used for the theoretical and empirical analyses of marriage markets.

Let $\tau_{x,y}$ be the transfer between $x$ and $y$ that occurs if they are matched.
The values that $x$ and $y$ obtain from the match are $u_{x,y}$ and $u_{y,x}$, respectively. Namely, $u_{x, y} = p_{x,y} + \epsilon_{x,y} -\tau_{x,y},\,\,\, u_{y,x} = p_{y,x} + \epsilon_{y,x} + \tau_{x,y}$,
where $\epsilon_{x,y}, \epsilon_{y,x}$ are the prediction errors of unilateral preferences $p_{x,y}, p_{y,x}$.
They also have the option of remaining ``unmatched'' and staying single.
The utilities for the remaining unmatched are $u_{x,0} = \epsilon_{x,0}, u_{y,0} = \epsilon_{y, 0}$, since we assume $p_{x,0} = p_{y,0} = \tau_{x,0} = \tau_{0,y} = 0$.
Given transfers $\tau_{x,y}$, each user chooses the other user who maximizes his/her value.
Let $\mu_{x,y}$ and $\mu_{y,x}$ be the probabilistic demand of $x \in X$ towards $y \in Y\cup\{0\}$ and that of $y \in Y$ toward $x \in X\cup\{0\}$, respectively:
\begin{equation*}
    \mu_{x,y} = \mathrm{P}_{\epsilon}\left(y = \argmax_{y' \in Y\cup\{0\}}u_{x, y'}\right),\,\,\, \mu_{y,x} = \mathrm{P}_{\epsilon}\left(x = \argmax_{x' \in X\cup\{0\}}u_{y, x'}\right).
\end{equation*}
As known in TU matching literature \citep{Choo2006-kx, Galichon2021-mq}, there exists a transfer $(\tau_{x,y})_{x\in X, y\in Y}$ such that the demands of both sides coincide $\mu_{x,y} = \mu_{y,x}$
for each $(x, y) \in X \times Y$ pair, and we call this $\mu$ the equilibrium matching.

To compute $\mu$, we assume that the distribution $F$ of errors $\epsilon_{x,y}, \epsilon_{y,x}$ is the standard type-I extreme value distribution (the standard Gumbel distribution), as in \citet{Choo2006-kx}.
Then, for each $x\in X, y \in Y$, we have
\begin{equation*}
    \mu_{x,y} = \mu_{y,x} = \exp\left(\frac{p_{x,y} + p_{y,x}}{2}\right)\sqrt{\mu_{x,0}}\sqrt{\mu_{y,0}},
\end{equation*}
where $\mu_{x,0}, \mu_{y,0}$ are the probabilities with which they choose to remain unmatched.
One can interpret the term $\exp\left(\frac{p_{x,y} + p_{y,x}}{2}\right)$ as a reciprocal preference in the context of RRSs and the term $\sqrt{\mu_{x,0}}\sqrt{\mu_{y,0}}$ as a factor that reflects the capacity of the users.
Combining this with the following constraint:
\begin{equation*}
    \sum_{y\in Y}\mu_{x,y} + \mu_{x,0} = 1 \ \ \forall x \in X, \ \ \mathrm{and} \ \ \sum_{x  \in X}\mu_{y,x} + \mu_{y, 0} = 1 \ \ \forall y \in Y,
\end{equation*}
we can derive the equilibrium matching by solving a convex optimization.

\section{Application and Challenges in Online Dating}
\label{sec:app}

\begin{wrapfigure}{r}{0.28\textwidth}
\centering
\includegraphics[width=\hsize]{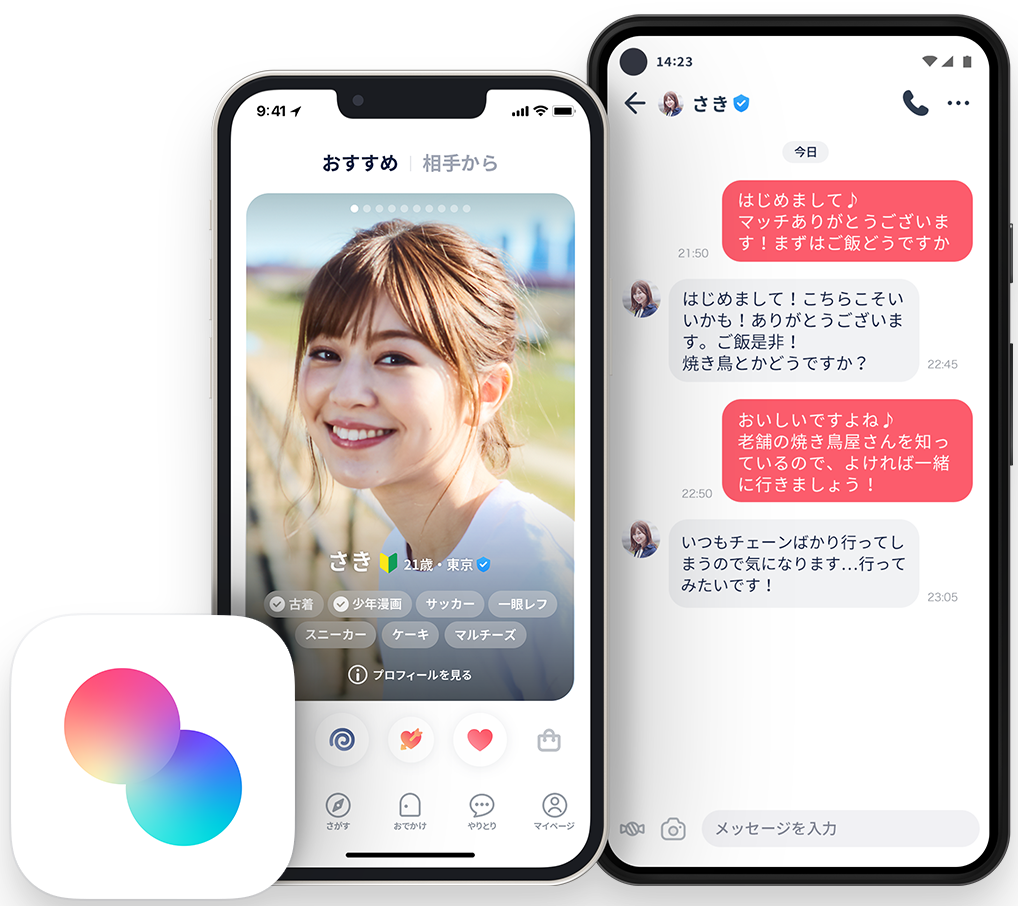}
\caption{The screenshots of tapple. Left: A recommended woman is shown for the user. The user can swipe to the right (like) or to the left (nope). Right: The user can chat with the matched partner.}
\label{fig:fig1}
\end{wrapfigure}

Tapple is a Japanese online dating platform that serves more than 7 million registered users.\footnote{\url{https://www.cyberagent.co.jp/news/detail/id=26472}.}
Once onboarded, users are shown the photos and profile information of recommended candidates~(Fig~\ref{fig:fig1}: Left).
They can send either ``like'' or ``nope'' to the recommended candidates. A candidate user who receives ``like'' can either ``thank'' for matching or ``sorry'' for rejecting. They are ``matched'' if the like recipient responds with "thank" and then allowed to chat (Fig~\ref{fig:fig1}: Right).

\begin{figure*}[t]
    \centering
    \includegraphics[keepaspectratio, width=0.9\linewidth]{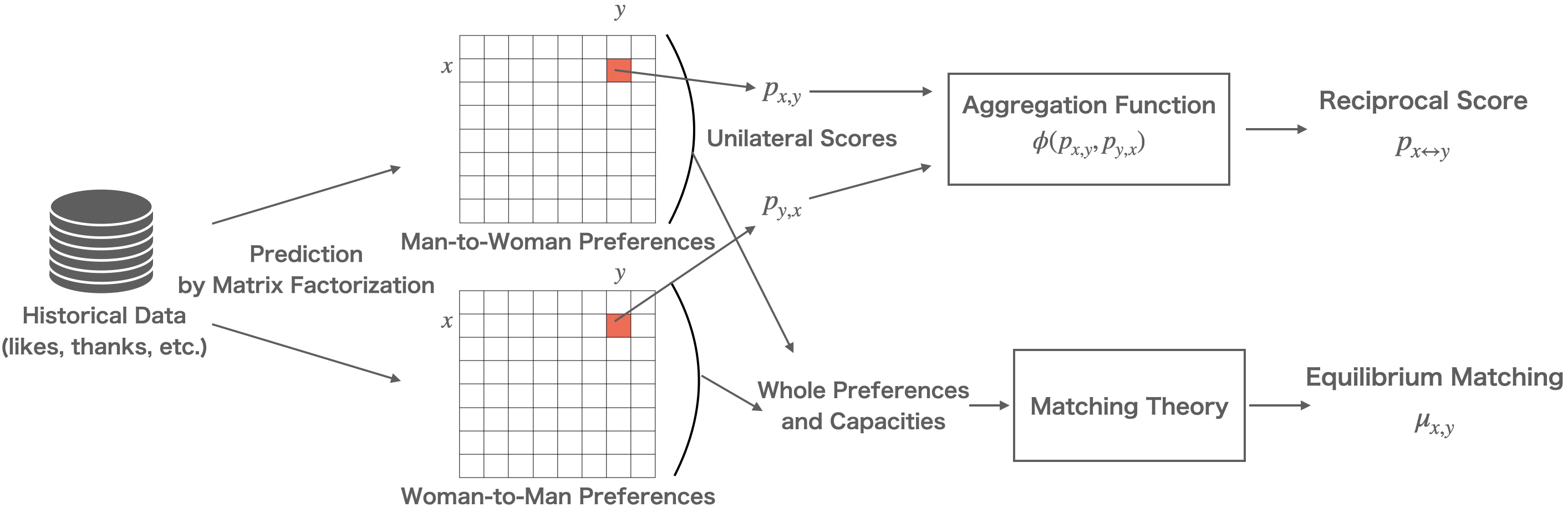}
    \caption{A visual description of the calculation of reciprocal scores. Details are given Section~\ref{sec:app}}
    \label{fig:fig2}
\end{figure*}

The application procedure is shown in Fig.~\ref{fig:fig2}.
Unilateral preference scores between men and women $p_{x,y}$ and the opposite $p_{y,x}$ are predicted by matrix factorization (MF) using unilateral historical feedback such as ``likes'' and ``thanks''.
When using a conventional RRS approach,
the following preference fusion step integrates unilateral scores into a reciprocal score $p_{x \leftrightarrow y}$.
In the prediction step for each man $x$ and each woman $y$,
the system sorts the candidate users according to $p_{x \leftrightarrow y}$ and recommends the users at the top of the ranked list.

Our project at tapple is to replace off-the-shelf fusion procedure with a cutting-edge MTRS.
To this end, we implemented a new TU-matching algorithm based on \citet{Choo2006-kx}.
Unlike standard RRSs, which only rely on preferences, our MTRS considers \emph{both} preferences and capacities by constructing a reciprocal score with transfer $\tau_{x, y}$.
Our MTRS thus mitigates the extreme concentration of ``likes'' and ``matches'' for enhancing overall user experience.

Scalability is a critical challenge for MTRSs, whereas this point has not been explored in depth.
To allow efficient estimation of equilibrium matching, we adopt the recently proposed iterative proportional fitting procedure (IPFP)~\citep{Decker2013-pj,Galichon2021-mq}.
We can obtain the optimal $\mu_{x,0}$ and $\mu_{y,0}$ by iteratively applying the following closed-form update formulas:
\begin{align*}
&\sqrt{\mu_{x,0}} \leftarrow \sqrt{1 - \left(\frac{1}{2}\sum_{y \in Y}\tilde{p}_{x \leftrightarrow y}\sqrt{\mu_{y,0}}\right)^2} - \frac{1}{2}\sum_{y \in Y}\tilde{p}_{x \leftrightarrow y}\sqrt{\mu_{y,0}},\\
&\sqrt{\mu_{y,0}} \leftarrow \sqrt{1 - \left(\frac{1}{2}\sum_{x \in X}\tilde{p}_{x \leftrightarrow y}\sqrt{\mu_{x,0}}\right)^2} - \frac{1}{2}\sum_{x \in X}\tilde{p}_{x \leftrightarrow y}\sqrt{\mu_{x,0}},
\end{align*}
where $\tilde{p}_{x \leftrightarrow y}=\exp\left(\frac{p_{x,y}+p_{y,x}}{2}\right)$.

Still, the above implementation does not meet the requirements in computational feasibility in tapple due to the million-scale users.
The dominant factor of complexity comes from the full computation of preference matrices and the iterative closed-form update of $\mu_{x,0}$ and $\mu_{y,0}$; this grows specifically to $O(|X||Y|(d+T))$ where $d$ is the dimensionality of MF, and $T$ is the number of iterations for IPFP.
This complexity can be a severe bottleneck; in the history of recommender systems, $O(|X||Y|)$ is prohibitively large and hence avoided, e.g. the computational cost of iALS~\citep{hu2008collaborative} for a single step is $O((Nd^2+(|X|+|Y|)d^3)T)$ where $N$ is the number of observed feedback, which is often small due to feedback sparsity (i.e., $N \ll |X|+|Y|$).
Maintaining the full score matrix $\{\tilde{p}_{x \leftrightarrow y} \mid x \in X, y \in Y\} \in \mathbb{R}^{|X| \times |Y|}$ is also space-consuming.
To alleviate this inefficiency issue, we explore approximated computation techniques based on locality sensitive hashing~\citep{gionis1999similarity} and approximate nearest neighbor search~\citep{wang2015learning}
to approximate the sum over all users or items in the update of $\sqrt{\mu_{x,0}}$ and $\sqrt{\mu_{y,0}}$, leading to the cost of $O(|X||Y|)$ for a single step.

\citet{Chen2021-oj} is the only study that examines TU matching for recommendation in online dating platform. 
They divide users into disjoint groups according to their characteristics and predict unilateral preferences between \emph{groups} using ordinary least squares (OLS).
While this grouping process may drastically reduce computational cost under coarse groups, it leads to identical recommendation results for users within the same group.
\footnote{Although \citet{Chen2021-oj} also conduct offline experimentation with MF, they randomly sample $1,000$ men and $305$ women for each experiment.}
On the contrary, our approach enables \emph{individual} matching due to the unilateral preferences based on MF while maintaining scalability by approximating the estimation of equilibrium matching.

\section{Future directions}

There are many other considerable future research directions.
\textbf{The Application of other Matching Algorithms}: In addition to TU matching, various matching algorithms are presented in non-transferable (NTU) matching, in which two types of agent are matched by a centralized authority and no monetary transfers occur between matched agents, such as the pioneering work of \citet{Gale1962-gc}.
Most NTU matching algorithms rely on the preference orderings of people and are therefore suitable to combine with learning-to-rank methods \citep{Cao2007-ga}.
\textbf{MTRSs meets algorithmic fairness}: There is a large literature on matching with various constraints, for example, matching algorithms with affirmative actions \citep{Hafalir2013-xg} and regional constraints \citep{Kamada2015-pe}.
Its application in MTRSs could be helpful for fairness issues in RRSs \citep{burke2017multisided}. \textbf{Bandit algorithms and MTRSs}: As in standard RSSs, it is also important in RRSs to balance exploration and exploitation.
In matching theory, matching with bandit algorithms in which agents learn their own preferences in the process is a growing literature~\citep{Jagadeesan2021-ke, Liu2020-ae}, and its application in RRS would be interesting future work.
\textbf{Online Experiment}: The experiment design should be carefully tailored for RRSs in the sense that users interact with each other through the platforms, which obviously violates SUTVA. Structural estimation should also aid in robust evaluation~\citep{nandy2021b,fong2020search,jung2021secret}.


\bibliographystyle{unsrtnat}
\bibliography{bib}

\end{document}